%% file: manuscript.tex
\documentclass[apsprl,twocolumn,bibnotes,superscriptaddress]{revtex4-2}

\input{source/preamble}

\defaultbibliographystyle{apsrev4-2}

\begin{document}

\input{source/main}

\input{source/supplements}
\end{document}

%% file: source/preamble.tex
\usepackage{graphicx,subfigure}
\usepackage{changes}
\usepackage{amsmath,amssymb}
\usepackage{mathtools}
\usepackage{multirow}
\usepackage{textcomp}
\usepackage{float}
\usepackage{xcolor}
\usepackage{titletoc}
\usepackage{ulem}
\usepackage{dsfont}
\usepackage{siunitx}
\DeclareSIUnit\gauss{G}
\usepackage{bibunits}
\usepackage{import}
\usepackage{braket}
\usepackage{nicefrac}
\usepackage{circledsteps}
\usepackage{comment}
\usepackage{bm}
\usepackage{bbm}
\usepackage[english]{babel}
\usepackage{hyperref}
\hypersetup{colorlinks=true,linkcolor=blue,citecolor=blue,breaklinks}

\newcommand{\micro}[1]{\ensuremath{\mu\mathrm{#1}}}

\renewcommand{\micro}[1]{\ensuremath \mu\mathrm{#1}}

\let\oldsfdefault\sfdefault
\usepackage[scaled=0.92]{helvet}
\renewcommand{\sfdefault}{\oldsfdefault}

\definechangesauthor[name=Immanuel Bloch, color=red]{ib}

\usepackage{layouts}

%% file: source/main.tex
\begin{bibunit}
\input{source/title}

\date{\today}

\begin{abstract}

Periodically driven quantum systems can realize novel phases of matter that are not present in time-independent Hamiltonians. One important application is the engineering of synthetic gauge fields, which opens the realm of topological many-body physics to neutral atom quantum simulators. 
In this work, we leverage a neutral atom quantum simulator to experimentally realize the strongly-interacting Mott-Meissner phase in large-scale, bosonic flux ladders with 48 sites at half filling. By combining quantum gas microscopy with local basis rotations, we reveal the emerging equilibrium particle currents with local resolution across large systems. We find chiral currents exhibiting a characteristic interaction scaling, providing direct experimental evidence of the interacting Mott-Meissner phase. Moreover, we benchmark density correlations with numerical simulations and find that the effective temperature of the system is on the order of the tunnel coupling.
Our results demonstrate the feasibility of scaling periodically driven quantum systems to large, strongly correlated phases, paving the way for exploring topological quantum matter with single-atom resolution and control.

\end{abstract}
\maketitle


Floquet engineering via periodic modulation has emerged as a powerful tool for Hamiltonian engineering~\cite{eckardt_colloquium_2017}. It enables the exploration of novel quantum matter across various platforms including photonics~\cite{ozawa_topological_2019}, neutral atoms in optical lattices~\cite{goldman_periodically_2014,bukov_universal_2015,eckardt_colloquium_2017}, superconducting qubits~\cite{roushan_chiral_2017,rosen_synthetic_2024} as well as Rydberg atom arrays~\cite{scholl_microwave_2022,zhao_floquet-tailored_2023}, and increasingly, also solid-state systems~\cite{mciver_light-induced_2020,weitz_lightwave-driven_2024}. A central goal of current research with synthetic quantum systems is to combine Floquet-engineered gauge fields~\cite{AidelsburgerArtificial2018,CooperTopological2019} with strong inter-particle interactions. Their interplay can give rise to topologically ordered phases of matter, both of fundamental interest and with practical applications for example in fault-tolerant quantum computing~\cite{kitaev_fault-tolerant_2003-1,nayak_non-abelian_2008}. Despite this potential, large-scale quantum simulations of such systems have been hindered by the susceptibility of interacting Floquet systems to heating~\cite{dalessio_long-time_2014, lazarides_equilibrium_2014, ponte_periodically_2015, thanasilp_quantum_2021, zhengEfficientlyExtractingMultiPoint2022, bilitewski_population_2015, bilitewskiScatteringTheoryFloquetBloch2015}, restricting experimental studies to few-particle systems~\cite{taiMicroscopyInteractingHarperHofstadter2017,ClarkObservation2020,leonardRealizationFractionalQuantum2023,wang_realization_2024}.

In this work, we experimentally realize the interacting Mott-Meissner phase in large bosonic ladder systems with a synthetic gauge field on a neutral atom quantum simulator. The gauge field breaks time-reversal symmetry, realizing the Hofstadter-Bose-Hubbard (HBH) model that is known to host topologically ordered phases such as fractional Chern insulators~\cite{hafeziFractionalQuantumHall2007,palmerOpticalLatticeQuantum2008,moller_composite_2009}. A quasi-1D ladder geometry is the minimal system to observe orbital physics, and hence, ideally suited to benchmark experimental implementations. Furthermore, interacting flux ladders were predicted theoretically to exhibit an extraordinarily rich phase diagram, featuring for example vortex and Meissner states, chiral superfluids and chiral Mott-insulators, or charge density wave states~\cite{piraudVortexMeissnerPhases2015,greschnerSpontaneousIncreaseMagnetic2015,greschnerSymmetrybrokenStatesSystem2016,didioPersistingMeissnerState2015,citroHallResponseInteracting2024,buserProbingHallVoltage2021,buserSnapshotbasedCharacterizationParticle2022}. However, to this date, optical flux ladders have only been realized with periodic driving either in the non-interacting or weakly-interacting limit~\cite{atalaObservationChiralCurrents2014,manciniObservation2015}, or via synthetic dimensions~\cite{stuhl_visualizing_2015,zhouObservation2023,zhouMeasuring2024,liang_chiral_2024,chenInteractiondrivenBreakdownAharonovBohm2024}.

\begin{figure}[!t]
    \centering
    \includegraphics[width=\columnwidth]{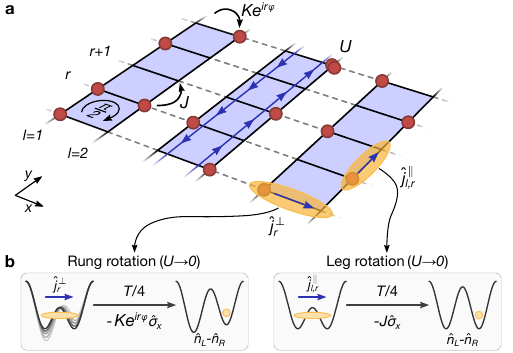}
    \caption{\textbf{Experimental setup and current measurement in a many-body system with complex tunneling.} \textbf{a}, Hofstadter-Bose-Hubbard model on two-leg ladders, parametrized by a real-valued tunnel coupling $J$ along the leg direction, complex-valued tunnel coupling $Ke^{ir\varphi}$ along the rung direction and a repulsive on-site interaction $U$. Here, $r$ is the rung index, and $l={1,2}$ indexes the leg. Persistent particle currents emerge due to the synthetic magnetic field (middle ladder, blue arrows). \textbf{b}, Measurement of currents in a many-body system with complex tunneling. After quenching the interactions to zero via a Feshbach resonance, the leg current operator is measured via a double well (DW) basis rotation with real coupling (right), while for the rung current operator it has to be performed in a driven DW with complex coupling (left). $\hat{\sigma}_x$ is the Pauli-\textit{X} operator and $T$ the tunneling time in the DW as defined in the main text.}
    \label{fig:experiment_setup}
\end{figure}

A crucial challenge for realizing large many-body phases consists in identifying a suitable adiabatic preparation sequence. In this work we present two different pathways and identify stable parameter regimes with minimal Floquet heating during the entire ramp to realize strongly-interacting flux ladders with 48 sites at half filling. To study the properties of our many-body system, we combine quantum gas microscopy with local basis rotations~\cite{impertroLocalReadoutControl2024} to measure particle currents with local resolution across large systems. To this end, we extend our previously developed technique to many-body systems and measurements of currents on bonds with complex tunnel coupling. We experimentally probe the local current distribution with full spatial resolution, revealing the emergence of equilibrium chiral currents that are directly connected to the topologically protected chiral edge modes in a 2D quantum Hall system~\cite{hugel_chiral_2014}. We further employ a Feshbach resonance to tune the interaction strength over a wide range and study its effect on the chiral currents. This reveals a characteristic interaction scaling, which is a direct experimental signature and hallmark feature of the interacting Mott-Meissner phase. Finally, we benchmark our experimental results against numerical simulations for small systems to estimate the effective temperature, which is on the order of the tunnel coupling. This establishes an important reference for future studies of topological quantum matter in periodically-driven many-body systems.

\textbf{Experimental setup}. -- The experiment is carried out in a cesium quantum gas microscope, which allows us to create strongly interacting quantum gases and probe occupations with single-site and currents with single-bond resolution~\cite{impertro_unsupervised_2023,wienandEmergenceFluctuatingHydrodynamics2024,impertroLocalReadoutControl2024}. We start with a \textsuperscript{133}Cs Bose-Einstein condensate in a single two-dimensional (2D) plane of a vertical optical lattice, which is then loaded into a 2D horizontal lattice geometry. The horizontal lattice is comprised of a bichromatic superlattice along $x$ (spacings $a_\mathrm{s} = \SI{383.5}{nm}$, $a_\mathrm{l} = \SI{767}{nm}$) as well as a monochromatic lattice (spacing $a_\mathrm{s}$) in the $y$-direction, realizing a chain of double wells along $x$ that are coupled in the perpendicular direction. In plane, the system is confined in a box potential projected via a digital micromirror device (DMD), with a size of $40\times40$ lattice sites. Using a deep long $x$-lattice and shallow short lattices, the dynamics is constrained to $20$ independent copies of isolated ladders with a length of up to $40$ sites (Fig.~\ref{fig:experiment_setup}a). To engineer an effective magnetic flux, we use a laser-assisted tunneling scheme based on an additional, superimposed optical running wave. In an effective Floquet description, this time-periodic modulation results in a complex-valued tunnel coupling along the rung direction of the form $Ke^{i\varphi(l,r)}$~\cite{aidelsburgerExperimentalRealizationStrong2011,atalaObservationChiralCurrents2014}. Here, $\varphi(l,r)$ is a complex phase factor, which varies in space and creates an artificial magnetic field [see Supplementary Material (SM)]. Its spatial dependence is determined by the lattice geometry and in our case is fixed to increase by $\pi/2$ per bond, resulting in a synthetic magnetic flux of $\pi/2$ per plaquette (a quarter flux quantum). In the presence of the complex tunnel coupling, the system can be described by the HBH model~\cite{harper_single_1955, hofstadter_energy_1976}
\begin{align}
    \hat{\mathcal{H}} = &\sum_{l,r} \left[ -J \left( \hat{a}^\dagger_{l,r+1}\hat{a}^{\phantom\dagger}_{l,r} +\mathrm{ h.c.} \right) + \frac{1}{2}U \hat{n}_{l,r} \left(\hat{n}_{l,r}-1\right) \right] \nonumber\\ &- K \sum_{r} \left( e^{i r \varphi} \hat{a}^\dagger_{2,r} \hat{a}^{\phantom\dagger}_{1,r} + \mathrm{h.c.} \right),
    \label{eq:hh_hamiltonian}
\end{align}
where $\hat{a}^\dagger_{l,r}$ and $\hat{n}_{l,r}$ are the bosonic creation and particle number operators for site $l=1,2$ of the $r$-th rung, $J$ and $K$ are the tunnel couplings in the leg and rung direction, respectively, $U$ is the on-site interaction strength and $\varphi=\pi/2$ denotes the Peierls phase. This Hamiltonian hosts a multitude of different ground state phases depending on filling, flux and coupling ratio $K/J$~\cite{piraudVortexMeissnerPhases2015, greschnerSpontaneousIncreaseMagnetic2015, greschnerSymmetrybrokenStatesSystem2016}. Characteristic for each phase are different configurations of persistent particle currents that emerge due to the synthetic magnetic field. In our work, we focus mainly on the \textit{Meissner} phase, which appears for large rung couplings $K/J>(K/J)_\mathrm{cr}$~\cite{piraudVortexMeissnerPhases2015}. It is characterized by homogeneous, chiral leg currents and vanishing rung currents (Fig.~\ref{fig:experiment_setup}a). For strong interactions, it has been predicted to be a fractional Mott-insulator at half filling, with a chiral current magnitude that depends characteristically on the interaction strength~\cite{piraudVortexMeissnerPhases2015}.

The operators describing such currents along the leg ($\hat{j}^\parallel_{l,r}$) and rung ($\hat{j}^\perp_{r}$) direction (see Fig.~\ref{fig:experiment_setup}a) are given by~\cite{kesslerSinglesiteresolvedMeasurementCurrent2014,piraudVortexMeissnerPhases2015}
\begin{align}
    \hat{j}^\parallel_{l,r} &= i J \left( \hat{a}^\dagger_{l,r+1} \hat{a}_{l,r} - \hat{a}^\dagger_{l,r} \hat{a}_{l,r+1} \right) \label{eq:leg_cur_op}\, \text{and} \\  
    \hat{j}^\perp_{r} &= i K \left( e^{-i r \varphi} \hat{a}^\dagger_{1,r} \hat{a}_{2,r} - e^{i r \varphi} \hat{a}^\dagger_{2,r} \hat{a}_{1,r}\right). \label{eq:rung_cur_op}
\end{align}

\begin{figure*}[t]
    \centering
    \includegraphics[width=\textwidth]{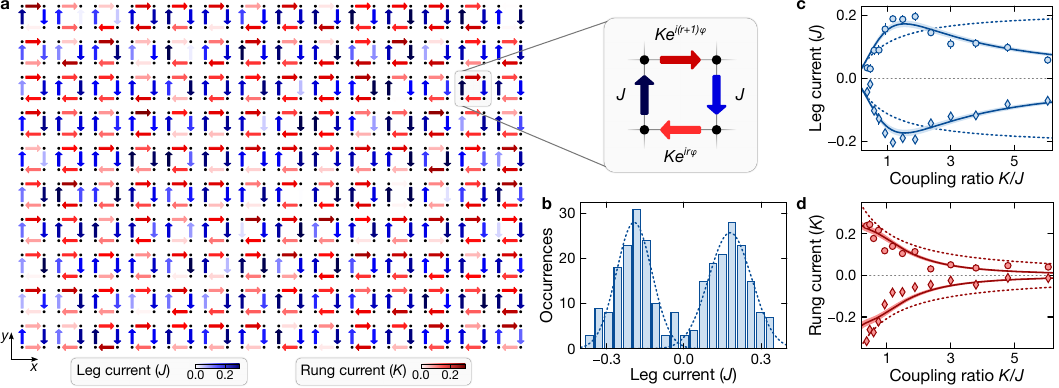}
    \caption{\textbf{Ground-state currents in isolated plaquettes with interactions.} \textbf{a}, Spatially-resolved map of the currents across a large array of 140 isolated plaquettes for $K/J\simeq1.4$ and $U/K\simeq10$. The direction of the current is indicated by the arrow, and the current magnitude is encoded in the color, where the leg currents are shaded in blue and the rung currents in red. Zoom-in: An exemplary plaquette, indicating the orientation of the real (complex) tunnel couplings on the leg (rung) bonds as defined in Eq.~(\ref{eq:hh_hamiltonian}). \textbf{b}, Distribution of the leg currents across the entire system shown in (a). The positive bonds have a mean current (1$\sigma$-deviation) of $0.18(8)\,J$ and the negative bonds $-0.19(8)\,J$, respectively, as illustrated by the normal distributions (dashed line). \textbf{c,d}, Scaling of the leg (c) and rung currents (d) as a function of $K/J$ analyzed over the region shown in (a). The solid line is a fit of an ED simulation of the ideal currents, with the amplitude as a single free parameter, yielding $0.78(4)$ for the legs and $0.71(4)$ for the rungs; the shaded area denotes the $1\sigma$-confidence interval of the fit. The dashed lines indicate the currents in a non-interacting plaquette with the same fit amplitude. The error bars denote the standard-error-of-the-mean (SEM), and if not visible, are smaller than the marker size. All numerical simulations take into account the reduced flux in isolated plaquettes of $0.71(2)\times\pi/2$ (details in the SM).}
    \label{fig:plaquettes}
\end{figure*}

Experimentally, we can measure the leg current [Eq.~(\ref{eq:leg_cur_op})] with single-bond resolution by locally rotating the measurement basis using an optical superlattice~\cite{kesslerSinglesiteresolvedMeasurementCurrent2014, impertroLocalReadoutControl2024}. As illustrated in Fig.~\ref{fig:experiment_setup}b (right panel), we project bonds locally into symmetrically coupled, isolated double wells (DW), by suddenly turning on an additional long-period lattice along the leg direction. The periodic time-evolution in each bond under the DW Hamiltonian can then be interpreted as a local rotation of the measurement basis. After holding for a quarter rotation period [$T/4=h/(8J_\mathrm{DW})$, where $J_\mathrm{DW}$ is the DW coupling and $h$ is Planck's constant], the local bond current is encoded in the density difference $\hat{n}_L-\hat{n}_R$ and can be directly read out using optical imaging. For a generic many-body state, interactions need to be switched off during the rotation, which otherwise modify the applied transformation. We implement this by switching the scattering length to approximately zero via a magnetic Feshbach resonance. To measure the rung currents [Eq.~(\ref{eq:rung_cur_op})], we have to adapt the protocol to account for the synthetic gauge field. Otherwise, the previous DW rotation would measure only the trivial, laser-induced phase $\varphi(l,r)$ rather than the emerging ground-state currents. To measure the actual rung current operator, we perform the basis rotation in the presence of the periodic drive, i.e., in DWs that have the same spatially-varying complex coupling phase as the rung bonds (Fig.~\ref{fig:experiment_setup}b, left panel, and SM). In summary, this technique provides access to snapshots of local particle currents on all bonds with microscopic resolution, which we will demonstrate in the following.

\textbf{Results}. -- We start by investigating the ground states of isolated plaquettes with two interacting particles each. This is an ideal system to benchmark the current detection in a many-body phase for both real and complex couplings, as it hosts stable currents that circulate around all four bonds of each plaquette~(Fig.~\ref{fig:plaquettes}a, zoom-in). To prepare the plaquette ground states, we begin with a product state where every 'rung' is occupied by one particle. Both long lattices in the $x$- and $y$-direction are kept deep throughout the sequence to define the plaquette geometry and suppress tunneling between the plaquettes (residual coupling in the leg direction $J'/h<\SI{5}{Hz}$, in the rung direction $K'/h=\SI{1.5}{Hz}$). In the presence of a strong tilt in the rung direction, the particle is initially localized in the energetically lower site. Next, we adiabatically turn on the running-wave modulation in $\SI{30}{ms}$ while simultaneously removing a weak additional tilt. This couples the two sites to a final strength $K/h=\SI{140(1)}{Hz}$ and delocalizes each particle symmetrically across a rung bond. In a final step, the short lattice in the leg direction is lowered in $\SI{15}{ms}$ to couple the two rungs and transfer the system to the plaquette ground state at a final $K/J$. Throughout the sequence, a strong repulsive on-site interaction is maintained. After finishing the state preparation, we measure the particle currents on all bonds by rotating the measurement basis as described above (Fig.~\ref{fig:experiment_setup}b).

Fig.~\ref{fig:plaquettes}a shows the distribution of currents across 140 isolated plaquettes, evaluated in a central sub-region of the whole system. We observe a homogeneous distribution of chiral currents originating from the homogeneous flux threading each plaquette. Analyzing the leg currents in more detail (Fig.~\ref{fig:plaquettes}b), we find that the width of the current distribution is approximately consistent with projection noise at the experimental sampling of 200 snapshots for each bond, with a slight broadening likely originating from on-site potential disorder (potential disorder amplitude $\sim h\times\SI{30}{Hz}$, see SM).

\begin{figure*}[t]
    \centering
    \includegraphics[width=\textwidth]{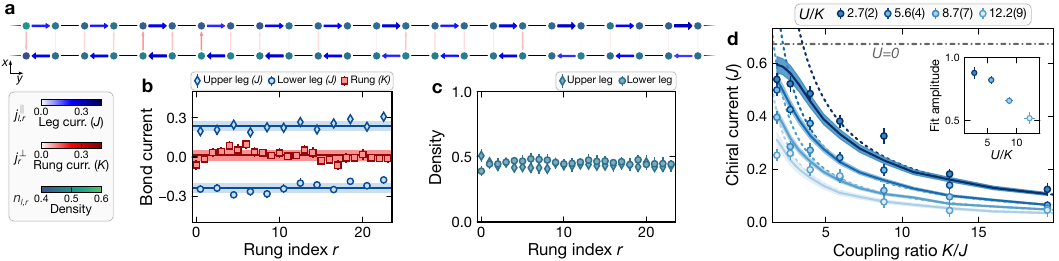}
    \caption{\textbf{Interacting ladders in the Meissner regime.} \textbf{a}, Spatially resolved density, leg and rung current distribution in the Meissner phase for $K/J=1.98(5)$ and $U/J=11.02(5)$. The width and the color of each arrow is given by the average magnitude of the respective bond current. Averaged over 140 repetitions and 14 ladders. \textbf{b}, Locally-resolved bond currents for the state in (a). The average currents are $0.24(4)\,J$ on the upper leg, $-0.23(3)\,J$ on the lower leg, and $0.01(4)\,K$ on the rungs, respectively, indicated also by the horizontal lines. \textbf{c}, On-site densities for the state in (a), yielding a homogeneous density profile with an average density of $0.45(2)$. \textbf{d}, Suppression of the chiral current with increasing interaction energy $U$ and coupling ratio $K/J$. The solid lines are fits of a DMRG simulation of the ideal chiral current with the amplitude as a single free parameter, the shaded areas denote the $1\sigma$-confidence interval of the fits, and the dashed lines are perturbative approximations using the effective spin-$1/2$ model, scaled to the same fit amplitude. The gray dot-dashed trace indicates the non-interacting current from an ED simulation at the same amplitude as the lowest $U/K$ measurement. The inset shows the fit amplitude as a function of $U/K$. The legend indicates the average $U/K$ for each curve, with the uncertainty denoting the $1\sigma$-variation throughout the $K/J$ range. Each data point is averaged over roughly 60 repetitions. In all plots, the error bars denote the SEM, and if not visible, are smaller than the marker size.}
    \label{fig:meissner_ladders}
\end{figure*}

To further study the ground-state phase diagram, we tune the coupling ratio $K/J$, and track the behavior of the bond currents. The interaction energy is on average $U/K\simeq9.8$, varying between $U/K=8.5(3)...11.0(3)$ as we tune $K/J$ via the short $y$-lattice depth. The measured dependency of the leg currents is shown in Fig.~\ref{fig:plaquettes}c. After an initial rise, it exhibits a maximum around $K/J\approx1.5$ as well as a suppression of the currents towards higher $K/J$. The suppression is characteristic for the interacting state and is markedly different from the case of non-interacting plaquettes (cf. dashed lines). This behavior is in excellent agreement with numerical simulations based on exact diagonalization (ED) of the two-particle plaquette ground state apart from an overall reduction of the ideal current to $78(4)\,\%$. This is repeated for the currents on the rungs (Fig.~\ref{fig:plaquettes}d), where we find similarly good agreement.
The measured current amplitudes are likely limited by the finite switching speed of our offset coils, causing a residual nonzero $U$ during the basis rotation, as well as a not fully adiabatic state preparation.
The above measurements demonstrate our capability to resolve both types of bond currents for interacting states with local resolution. We note that the single-shot sampling of the current operators also allows to measure current-current correlation functions, which further reveal strong features due to the micromotion as detailed in the SM.

Next, we study extended ladder systems at half filling with tunable interactions. Realizing such a system with a Floquet scheme is highly non-trivial, since in addition to the challenges of an adiabatic preparation, drive-induced heating needs to be minimized. In particular, heating resonances have to be avoided during preparation, limiting the accessible parameter space~\cite{sun_optimal_2020}. To address this, we conduct extensive loss spectroscopy, identifying a narrow window around modulation frequencies of $\SI{5}{kHz}$ with negligible losses (atom loss $\sim 2\,\%$ compared to initial state), bounded from below by interaction-mediated heating and from above by interband resonances.

To prepare the Mott-Meissner phase we employ a \textit{rung coupling} sequence. We start with a product state of one particle per rung which is delocalized across both sites in the presence of the complex tunnel coupling. For strong interactions, this state corresponds already to a strongly-interacting Meissner-like state in the $(K/J\to\infty)$ limit, and is thus readily connected to the Meissner phase by increasing the leg coupling. After the state preparation, we probe the current and density distribution with local resolution. Fig.~\ref{fig:meissner_ladders}a shows a current and density map for a strongly-interacting Meissner state. We restrict the evaluation to the central region of the whole $40\times40$ site system to avoid edge effects due to the finite wall sharpness of the box potential, and average over all ladder copies (residual coupling between ladders $K'/h=\SI{1.5}{Hz}$). Note that we only access every other bond in the leg direction due to the DW-array created with a superlattice. We find strong, chiral currents along the leg bonds, uniformly distributed across the ladder, accompanied by strongly suppressed currents on the rungs (Fig.~\ref{fig:meissner_ladders}b), as it is characteristic for the Meissner phase. In addition, we find a homogeneous filling of on average $0.45(2)$ across the ladder without any imbalance between the legs, where the slight deviation from ideal half filling originates mostly from an imperfect initial state (Fig.~\ref{fig:meissner_ladders}c).

A second key feature arising from the strongly correlated nature of the state is a characteristic suppression of the chiral current with increasing interaction strength $U$ as well as coupling ratio $K/J$~\cite{piraudVortexMeissnerPhases2015}. This can also be understood by noticing that the Meissner ladder in the limit of $K\gg J$ and $U\gg J$ can be mapped onto a pseudo-spin-$\nicefrac{1}{2}$ chain, where the ground state in the $U\to\infty$ limit is a product state of rung triplets, i.e., each rung $r$ is in the state $\ket{\downarrow}_r=(\ket{1,0}_r+e^{ir\pi/2}\ket{0,1}_r)/\sqrt{2}$. At half filling, a perturbative relation for the chiral current, defined as the average difference between the two leg currents, ${j_\mathrm{c}=\frac{1}{2L}\left| \sum_r \langle \hat{j}^\parallel_{1,r}\rangle-\langle \hat{j}^\parallel_{2,r} \rangle \right|}$ can be derived as $j_\mathrm{c} = [J^2(4K+U)^2]/[2KU(2K+U)]$~\cite{piraudVortexMeissnerPhases2015}. We can experimentally verify this behavior by tuning the interaction strength independent from the tunnel couplings via a magnetic Feshbach resonance. Again using the rung coupling sequence, we set a fixed $K/h=\SI{155(1)}{Hz}$, and vary the final $J$ to tune $K/J$ for different interaction strengths. Fig.~\ref{fig:meissner_ladders}d shows the dependence of the chiral current $j_\mathrm{c}$ on the coupling ratio $K/J$ for four different average interaction energies. Note that $U$ varies slightly around the average value as a function of $K/J$ (calibration in the SM). Qualitatively, we find that the current is suppressed both for higher coupling ratio as well as with increasing interaction energy. This is in stark contrast to a non-interacting ladder, where the chiral current remains constant with $K/J$ (cf. gray dot-dashed trace in Fig.~\ref{fig:meissner_ladders}d). Comparing this more closely to theory, we observe good agreement of the scaling with a zero temperature DMRG simulation (solid lines), as well as the perturbative approximation for $K/J\gtrsim 5$. For low $U/K$, we observe up to $88(5)\,\%$ of the ideally predicted current strength, which then drops towards higher $U$ values. In agreement with the plaquette benchmark, the current amplitude is likely limited by a residual nonzero $U$ during the basis rotation, with an increasing effect for larger initial $U$.

\begin{figure}[t]
    \centering
    \includegraphics[width=\columnwidth]{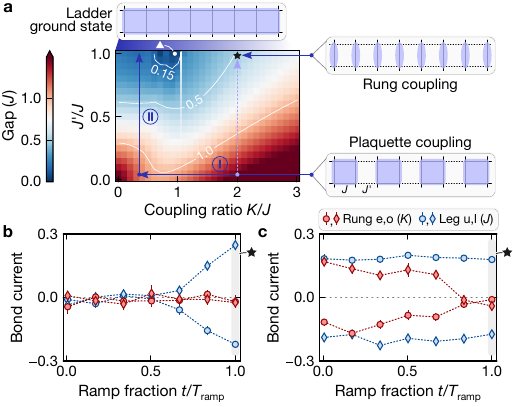}
    \caption{\textbf{Adiabatic preparation sequences.} \textbf{a}, Many-body gap as a function of coupling ratio $K/J$ and inter-plaquette coupling $J'/J$ for $U/J=10$, simulated using DMRG. The solid purple arrows and the illustrations indicate the paths taken by the rung and plaquette coupling sequences. The white triangle marks the gap closing, and the black star denotes the example Meissner state prepared below. \textbf{b},\textbf{c}, Evolution of the currents during the adiabatic ramp with duration $T_{\text{ramp}}$ in the rung coupling (b) and in the plaquette coupling (c) sequences for a final Meissner state with $K/J=1.98(5)$ and $U/J=11.02(5)$. The dashed lines are guides to the eye. u(l) denotes the upper(lower) leg; and e(o) indexes even(odd) rungs in an alternating fashion. The error bars denote the SEM, and if not visible, are smaller than the marker size. Each data point is averaged over roughly 30 repetitions and 14 ladders.}
    \label{fig:ladder_prep}
\end{figure}

While the previously used sequence allows for an efficient realization of Meissner states, it cannot be used to access the entire ground-state phase diagram of the ladder-HBH model. In particular, decreasing the coupling ratio $K/J$ below a critical value $(K/J)_\mathrm{cr}\approx 1$ that depends on the interaction strength, the system undergoes a phase transition into a \textit{vortex phase}. Here, the chiral Meissner current breaks up into several smaller current loops that are separated by current vortices. This results in a spatial modulation of the leg current alongside a decrease of the chiral current as well as nonzero rung currents. At the phase transition, a many-body gap closing separates the two phases, preventing a use of the rung coupling sequence for an adiabatic preparation (Fig.~\ref{fig:ladder_prep}a). To reach the vortex regime, we introduce a second tuning parameter $J'$, corresponding to a staggered tunnel coupling along the leg direction (vertical axis in Fig.~\ref{fig:ladder_prep}a). Making use of this additional parameter, the \textit{plaquette coupling} sequence starts from isolated plaquette ground states ($J'\ll J$), where we can prepare any $K/J$ adiabatically (Fig.~\ref{fig:ladder_prep}a, horizontal path \Circled{\small\textsf{I}}). In a second step, the plaquettes are connected together by increasing $J'/J\to1$ at constant $K/J$ (Fig.~\ref{fig:ladder_prep}a, vertical path \Circled{\small\textsf{II}}). To compare the two sequences, we track the evolution of the currents during the final ramps when preparing an interacting Meissner state. In the rung coupling sequence, we start with no currents on leg or rung bonds, and only the leg currents gradually build up with opposing sign (Fig.~\ref{fig:ladder_prep}b). In contrast to that, the plaquette sequence starts with initial opposing currents both on leg and rung bonds (Fig.~\ref{fig:ladder_prep}c). Upon connecting the plaquettes to a Meissner ladder, the leg currents remain finite, while the rung currents vanish. Both sequences can be employed to prepare Meissner states, but the plaquette coupling sequence results in slightly smaller currents due to the longer preparation path.

\begin{figure}[t]
    \centering
    \includegraphics[width=\columnwidth]{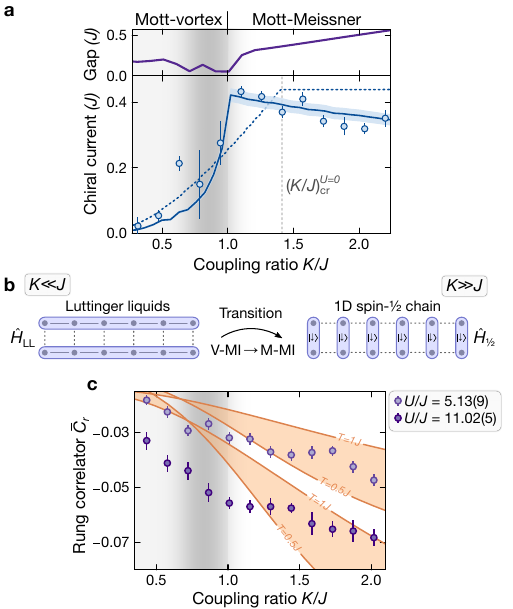}
    \caption{\textbf{Chiral current and density correlations across the phase diagram.} \textbf{a}, Average chiral current as a function of $K/J$ for $U/J=11.02(5)$ and $J/h=\SI{71(1)}{Hz}$, prepared using the plaquette coupling sequence. The solid line is a fit of the expected chiral current from a DMRG simulation with a scaling factor $A$ as a free parameter, yielding $A=0.57(3)$. The blue shaded area denotes the $1\sigma$-confidence interval of the fit. The dashed trace shows the non-interacting current, scaled down to the same amplitude, and the vertical line denotes the critical point without interactions. The upper panel indicates the many-body gap across the phase diagram. Each data point is averaged over roughly 80 repetitions. \textbf{b}, Effective description of the flux ladder system in terms of two coupled Luttinger liquids ($K\ll J$) and a 1D spin chain ($K\gg J, U\gg J$). \textbf{c}, Enhancement of the average rung-wise density anti-correlations with increasing interaction energy $U$ and coupling ratio $K/J$. The orange shaded areas indicate finite-temperature exact diagonalization simulations ($2\times6$ sites) of the density correlations from $k_\mathrm{B}T=0.5\,J$ (lower line) to $k_\mathrm{B}T=1\,J$ (upper line) for both interaction energies. Each data point was averaged over roughly 30 repetitions. The error bars denote the SEM, and if not visible, are smaller than the marker size.}
    \label{fig:ladder_phase_diagram}
\end{figure}

We use the plaquette coupling sequence to explore the ground-state phase diagram between vortex and Meissner regimes. One striking signature of the phase transition to the vortex phase is the sudden drop of the chiral current upon crossing the critical point at $(K/J)_\mathrm{cr}\approx1$. With strong interactions, the transition point is predicted to be significantly lower than in the non-interacting case, where it was previously shown that $(K/J)_\mathrm{cr}^{U=0}=\sqrt{2}$~\cite{orignacMeissnerEffectBosonic2001,atalaObservationChiralCurrents2014}. We experimentally probe this behavior by varying the coupling ratio $K/J$ for strong interactions. Fig.~\ref{fig:ladder_phase_diagram}a shows the experimentally measured chiral current around the phase transition between vortex and Meissner regimes. Indeed, we see a sudden drop of the chiral current around $(K/J)_\mathrm{cr}\approx1$, signaling a transition to the vortex regime. The observed behavior agrees well with a zero temperature DMRG simulation (solid line), with our measurements finding around $57(3)\,\%$ of the ideally predicted current. Compared to the rung coupling sequence, the current amplitude is reduced due to the longer preparation path (see SM for current lifetimes), as well as smaller tunnel couplings. Note that below the phase transition we observe enhanced fluctuations in the measured currents as reflected by the large error bars. We attribute this to a small many-body body gap, which makes the system highly susceptible to technical heating.

Lastly, we probe the fractional Mott-insulating nature of the Meissner phase. In the strong rung coupling limit (i.e., deep in the Meissner regime), the effective pseudo-spin-$\nicefrac{1}{2}$ model predicts the ground state to be a product state of rung triplets (Fig.~\ref{fig:ladder_phase_diagram}b). Adding a second particle to one rung costs an energy $\sim K$, similar to having a 1D Mott-insulating chain of these triplets with a fractional charge of $\nicefrac{1}{2}$ (also called \textit{rung-Mott insulator})~\cite{crepinPhaseDiagramHardcore2011,piraudVortexMeissnerPhases2015}. The presence of exactly one particle on each rung results in a strong density anti-correlation across the rungs, which is experimentally accessible. Fig.~\ref{fig:ladder_phase_diagram}c shows the measured rung density correlator $C_{r} = \langle \hat{n}_{1,r} \hat{n}_{2,r} \rangle-\langle \hat{n}_{1,r} \rangle\langle \hat{n}_{2,r} \rangle$, averaged over all rungs, as a function of $K/J$ for two different interaction energies. We find significantly negative density correlations, which are enhanced both for increasing $U$ as well as increasing $K/J$, in accordance with the prediction of the effective spin model. In the $U\to\infty$ and $K\gg J$ limit, one would expect a correlation of $-\nicefrac{1}{4}$, with the mass gap persisting as long as $K>J$. In comparison, the measured correlations are weakened due to the deviation from half-filling (where one strictly does not expect a perfect Mott state), finite $U$ and, most importantly, a finite temperature that softens the Mott-transition and decreases the anti-correlation.

In fact, the density correlations are very temperature-sensitive, which allows us to provide a rough estimate of the effective temperature of our many-body state. To this end, we perform small-scale ED simulations on $2\times6$ sites, revealing that rung-wise density correlations decay smoothly as the temperature increases from zero to the scale of the leg tunnel coupling. The simulation also accounts for initial state imperfections as well as parity projection. As shown by the orange shaded areas in Fig.~\ref{fig:ladder_phase_diagram}c, a comparison of the correlator strength with the simulation indicates a temperature on the order of $k_\mathrm{B} T \sim J$ in the Meissner phase, consistent also with the observed chiral current magnitudes. A temperature on this order is also compatible with the predicted elementary excitations with energy $K$ of the spin model, since at this scale doubly-occupied or empty rungs can be formed, both bringing the correlator closer to zero. In the vortex regime ($K<J$), the effective temperature is likely higher due to the smaller gap and overall lower energy scales. To predict finite-$T$ effects deep in the vortex phase, an effective description in terms of two weakly coupled Luttinger liquids can be applied (Fig.~\ref{fig:ladder_phase_diagram}b)~\cite{orignacMeissnerEffectBosonic2001,crepinPhaseDiagramHardcore2011,piraudVortexMeissnerPhases2015}. This shows that the Mott-gap is exponentially small in $K$, and furthermore the gapless excitations along the leg direction quickly wash out any current modulation or rung current patterns, rendering a direct observation of vortices challenging at a finite temperature (cf. simulated currents in the SM).

\textbf{Discussion}. -- In our work, we demonstrated an experimental realization of the interacting Mott-Meissner phase on large bosonic flux ladders with 48 sites at half filling. By combining quantum gas microscopy with local basis rotations, we uncovered the key microscopic features of this phase: persistent chiral currents along the legs of the ladder accompanied by strongly suppressed rung currents, a characteristic interaction-induced suppression of the chiral current, as well as a density anti-correlation across the rung bonds, which is a direct signature of the fractional-Mott-insulating behavior at half filling. Comparing our measurements with small-system numerics allowed us to estimate the effective temperature, setting a new benchmark for interacting, periodically-driven quantum systems, which provides an important reference point for future theoretical and experimental efforts.

The control and microscopic detection over a large, interacting ladder system, as shown here, opens an exciting avenue towards the realization of topological quantum matter. In particular, the platform can be directly employed to study transport phenomena and non-equilibrium dynamics in time-reversal-symmetry-broken many-body phases in novel ways using local current measurements~\cite{buserSnapshotbasedCharacterizationParticle2022, qiao_quench_2021, giri_flux-enhanced_2024,jian_defect_2023,citro_spectral_2020,citroHallResponseInteracting2024,huang_spatial_2024}. By further mapping out the rich phase diagram of interacting two-leg ladders, we can uncover various additional many-body phases, including vortex, chiral Mott/superfluid, and charge density wave states~\cite{piraudVortexMeissnerPhases2015,greschnerSymmetrybrokenStatesSystem2016}. Key future steps to this end involve developing advanced preparation techniques~\cite{wang_robust_2021,schindler_counterdiabatic_2024} and mitigating Floquet heating~\cite{viebahn_suppressing_2021}. Finally, extending this system via multi-leg ladders to a full 2D geometry provides a controlled pathway towards large-scale, analog quantum simulation of fractional Chern insulators with hundreds of atoms~\cite{palmSnapshotbasedDetection$ensuremathnufrac12$2022,wangMeasurableSignaturesBosonic2022,palmAbsenceGaplessMajorana2024}.

\vspace{2em}
\begin{center}
\textbf{ACKNOWLEDGEMENTS}
\end{center}
\vspace{0.5em}

The authors would like to acknowledge insightful discussions with Botao Wang, Paul Schindler, André Eckardt, Marcello Dalmonte, Eugene Demler, Fabian Grusdt, Matteo Rizzi, Alessio Recati and Wolfgang Ketterle. We further thank Christian Schweizer for early contributions to the experiment and Irene Prieto Rodríguez for valuable feedback on the manuscript. We received funding from the Deutsche Forschungsgemeinschaft (DFG, German Research Foundation) via Research Unit FOR5522 under project number 499180199, via Research Unit FOR2414 under
project number 277974659 and under Germany’s Excellence Strategy – EXC-2111 – 390814868 and from the German Federal Ministry of Education and Research via the funding program quantum technologies – from basic research to market (contract number 13N15895 FermiQP). This publication has further received funding under Horizon Europe programme HORIZON-CL4-2022-QUANTUM-02-SGA via the project 101113690 (PASQuanS2.1). 
S.H. was supported by the education and training program of the Quantum Information Research Support Center, funded through the National research foundation of Korea (NRF) by the Ministry of science and ICT (MSIT) of the Korean government (No.2021M3H3A1036573).

\newpage
\vspace{2em}
\begin{center}
\textbf{REFERENCES}
\end{center}
\vspace{0.5em}

\putbib[manuscript]
\end{bibunit}


\clearpage

%% file: source/title.tex
\title{Realization of strongly-interacting Meissner phases in large bosonic flux ladders}

\author{Alexander Impertro}\email{a.impertro@lmu.de}
\author{SeungJung~Huh}
\author{Simon Karch}
\author{Julian~F.~Wienand}
\author{Immanuel Bloch}
\author{Monika Aidelsburger}\email{monika.aidelsburger@physik.uni-muenchen.de}
    \affiliation{Fakult\"{a}t f\"{u}r Physik, Ludwig-Maximilians-Universit\"{a}t, 80799 Munich, Germany}
    \affiliation{Max-Planck-Institut f\"{u}r Quantenoptik, 85748 Garching, Germany}
    \affiliation{Munich Center for Quantum Science and Technology (MCQST), 80799 Munich, Germany}

%% file: source/supplements.tex
\begin{bibunit}
\setcounter{section}{0}
\setcounter{equation}{0}
\setcounter{figure}{0}
\setcounter{table}{0}
\renewcommand{\theequation}{S\arabic{equation}}
\renewcommand{\theHequation}{S\arabic{equation}}
\renewcommand{\thefigure}{S\arabic{figure}}
\renewcommand{\theHfigure}{S\arabic{figure}}
\renewcommand{\thetable}{S\arabic{table}}
\renewcommand{\theHtable}{S\arabic{table}}
\setcounter{page}{1}
\setcounter{affil}{0}

\author{Alexander Impertro}
\author{SeungJung~Huh}
\author{Simon Karch}
\author{Julian~F.~Wienand}
\author{Immanuel Bloch}
\author{Monika Aidelsburger}
    \affiliation{Fakult\"{a}t f\"{u}r Physik, Ludwig-Maximilians-Universit\"{a}t, 80799 Munich, Germany}
    \affiliation{Max-Planck-Institut f\"{u}r Quantenoptik, 85748 Garching, Germany}
    \affiliation{Munich Center for Quantum Science and Technology (MCQST), 80799 Munich, Germany}

\title{Supplementary Information for: \\ Realization of strongly-interacting Meissner phases in large bosonic flux ladders} 

\maketitle

\tableofcontents


\section{Experimental sequence}

\begin{figure*}[t]
    \centering
    \includegraphics[width=\textwidth]{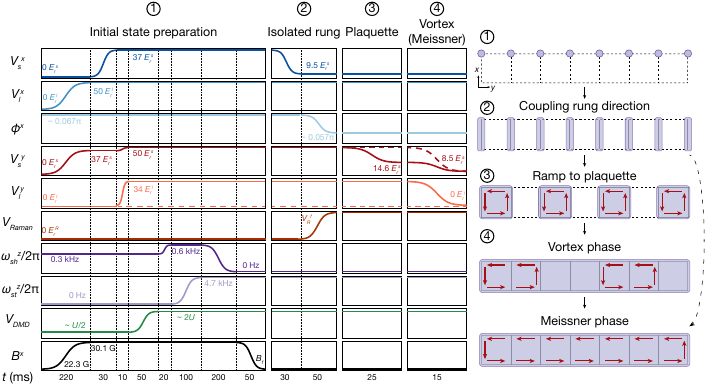}
    \caption{\textbf{Experimental sequence.} Full experimental sequence for initial state preparation and adiabatic preparation of the ground state phases. The solid (dotted) line is the preparation scheme for the vortex (Meissner) phase.}
    \label{fig:exp_seq}
\end{figure*}

In this section, we describe our experimental sequence to prepare the half-filling initial state, engineer an artificial magnetic field, and prepare ground states of the Hofstadter-Bose-Hubbard (HBH) Hamiltonian in a quasi-1D ladder geometry (Fig.~\ref{fig:exp_seq}).

\subsection{Initial state preparation}
Each experimental realization starts by preparing a Bose-Einstein condensate of $^{133}\rm{Cs}$ atoms in a single plane of a vertical lattice with $\SI{8}{\micro\meter}$ spacing (large-spacing vertical lattice). Horizontal confinement is provided by a repulsive box potential shaped by a digital micromirror device (DMD) illuminated with blue-detuned, incoherent light from a multi-mode laser diode ($\lambda=\SI{525}{nm}$). We then perform forced optical evaporation cooling by lowering the vertical lattice depth to lower the sample temperature further.

Next, we prepare a unity-filling Mott insulating state by loading the atoms into a 2D in-plane lattice geometry, consisting of a long-period lattice along the $x$- and a short-period lattice along the $y$-direction. The lattices are ramped in $\SI{220}{ms}$ to respective lattice depths of $37\,E_\mathrm{r,l}$ ($x$) and $50\,E_\mathrm{r,s}$ ($y$). Here, $E_\mathrm{r,s(l)} = h^2/8ma_\mathrm{s(l)}^2$ is the recoil energy of the short-period (long-period) lattice, $h$ is Planck's constant, $m$ is the atomic mass of cesium, and $a_\mathrm{s} = \SI{383.5}{nm}$ ($a_\mathrm{l} = 2a_\mathrm{s}$) denotes the short-period (long-period) lattice constant. During the lattice ramp, we simultaneously increase the scattering length from $270\,a_0$ to $540\,a_0$ by ramping the magnetic field to $\SI{30.1}{G}$. To realize tilted double wells (DW) along the $x$- direction, we superimpose an additional short-period lattice. The resulting superlattice potential is described by
\begin{align}
    V(x) = V_\mathrm{s} \cos^2\left(\pi x/a_\mathrm{s}\right) + V_\mathrm{l} \cos^2\left(\pi x /a_\mathrm{l} + \phi/2\right),
\end{align}
with $V_\mathrm{s(l)}$ being the potential depth of the short(long)-period lattice, and $\phi$ is the superlattice phase, which is stabilized via a frequency-offset lock as described in Ref.~\cite{impertroLocalReadoutControl2024}. The short-period lattice is increased in $\SI{30}{ms}$ to $50\,E_\mathrm{r,s}$ at a superlattice phase of $\sim 0.067 \pi$ . This tilt causes all atoms to be localized in the lower wells. Our initial state has a typical filling of $93(1)\,\%$ in the occupied leg and $0.5(3)\,\%$ in the empty leg, which is close to a half-filled Mott insulating state in a two-leg ladder lattice.

To enhance on-site interaction energies, we increase the vertical confinement by transferring the atoms into a vertical lattice with a smaller spacing of $\SI{1.06}{\micro\meter}$ (small-spacing vertical lattice). The vertical lattice is formed by the interference of two $\SI{1064}{nm}$ laser beams intersecting at an angle of $60^{\circ}$. A ring piezo actuator is used to control the vertical lattice phase by adjusting the path length difference between the two beams. We increase the depth of the large spacing vertical lattice over $\SI{20}{ms}$ to squeeze the wave function in the $z$-direction, and subsequently ramp up the small-spacing vertical lattice over $\SI{100}{ms}$ to a final trap frequency of $2\pi\times\SI{4.7}{kHz}$ along the $z$-direction (in-plane harmonic curvature of $2\pi\times\SI{20}{Hz}$). By optimizing the vertical lattice phase, we ensure that the atoms are loaded into a single plane of the small-spacing vertical lattice. Finally, the large-spacing vertical lattice is adiabatically switched off within $\SI{200}{ms}$.

Further information about the experimental setup, including details on the optical superlattices as well as their phase stabilization, can be found in Refs.~\cite{impertro_unsupervised_2023,wienandEmergenceFluctuatingHydrodynamics2024,impertroLocalReadoutControl2024}.

\subsection{Laser-assisted tunneling}
An artificial magnetic field with a total flux per plaquette $\varphi/2\pi = 1/4$ is generated via a laser-assisted tunneling scheme. As described in Refs.~\cite{aidelsburgerExperimentalRealizationStrong2011,impertroLocalReadoutControl2024}, we superimpose two nearly perpendicular laser beams with a wavelength of $\SI{1534}{nm}$ on the superlattice potential. To inhibit bare tunneling along the rung direction ($x$), the superlattice phase is set to $\phi_\mathrm{sl}\simeq0.057\pi$, introducing a large energy offset, $\Delta \gg J$. In detail, the energy offset is $\Delta/h=\SI{4.37}{kHz}$ compared to a bare tunnel coupling of $J/h=\SI{810}{Hz}$. The frequency difference between the two $\SI{1534}{nm}$ lasers is then set to satisfy the resonance condition, $\omega_1-\omega_2 = \Delta/\hbar$, enabling the realization of tunnel coupling with a spatially dependent phase.

\subsection{Ground state preparation}
\label{sec:sm_gs_prep}
Our protocol for preparing the ground state in the two-leg ladder HBH model with $n = 0.5$ and $\varphi/2\pi = 1/4$ relies on an adiabatic path that avoids a many-body band gap closing (Fig.~\ref{fig:exp_seq}, cf. also many-body gap diagram in Fig.~\ref{fig:ladder_prep} of the main text). Initially, the system is prepared in a half-filling product state, where the leg of the ladder with lower potential is occupied with one atom per site, while the higher leg remains empty. Subsequently, a running-wave modulation scheme is used to adiabatically increase the complex tunneling coupling $K$ along the rung direction at a constant flux. To prepare the Meissner phase, we adiabatically ramp the coupling along the leg direction $J$ to a final state with a coupling ratio above the critical point $K/J > (K/J)_\mathrm{cr}$. We also refer to this sequence as the \textit{rung coupling sequence}, as we start with isolated rungs that are subsequently coupled.

The same protocol cannot be applied to the vortex phase since this adiabatic path crosses the gap closing point at around $K/J \approx 1$. As an alternative, we develop a \textit{plaquette coupling sequence} that prepares the ground state of isolated plaquettes and then connects them together by ramping the inter-plaquette coupling strength. Introducing a different intra- and inter-plaquette coupling $J$ and $J'$ maintains a finite gap along the entire dynamical evolution, which allows us to explore the whole phase diagram as a function of $K/J$ while maintaining near adiabaticity. In a final step, all plaquettes in the ladder are connected by ramping $J'/J \rightarrow 1$ to prepare the final ladder ground state, including states in the vortex regime.

\subsection{Current measurement}
The particle current is defined using the continuity equation, 
\begin{equation}
    \frac{d}{dt}\hat{n}_{l,r} + \sum_{\langle l',r' \rangle} \hat{j}_{l,r \rightarrow l',r'} = 0,
\end{equation}
where $\hat{n}_{l,r}$ is the particle number operator for site $l=1,2$ of the $r$-th rung, $\hat{j}_{l,r \rightarrow l',r'}$ is the particle current operator from the site $l,r$ to a nearest-neighboring site $l',r'$~\cite{kesslerSinglesiteresolvedMeasurementCurrent2014,piraudVortexMeissnerPhases2015}. In the HBH model in a two-leg ladder, the particle current can be obtained using the Heisenberg equation of motion
\begin{multline}
    \frac{d}{dt}\hat{n}_{l,r} = -iJ(\hat{a}_{l,r+1}^{\dagger}\hat{a}^{\phantom\dagger}_{l,r}-\hat{a}_{l,r}^{\dagger}\hat{a}^{\phantom\dagger}_{l,r+1}) \\ -iK(e^{-ir\phi}\hat{a}_{1,r}^{\dagger}\hat{a}^{\phantom\dagger}_{2,r}-e^{ir\phi}\hat{a}_{2,r}^{\dagger}\hat{a}^{\phantom\dagger}_{1,r})
\end{multline}
where $\hat{a}_{l,r}^{\dagger}$ is the bosonic creation operator for site $l=1,2$ of the $r$-th rung, and $J$ and $K$ are the tunnel coupling strengths along the leg and rung direction, respectively. The particle current has two contributions: the first term corresponds to the current along the leg $\hat{j}_{l,r}^\parallel$, while the second term is the current along the rung direction $\hat{j}_{l,r}^\perp$.

To reveal the particle current with single bond resolution, we rotate the measurement basis from density to current using double well manipulations prior to single-site imaging~\cite{impertroLocalReadoutControl2024}. To measure the leg current operator, the system is projected onto isolated double wells along the leg direction, here denoted by the sites $(l,r)$ and $(l,r+1)$. We then increase the tunneling coupling strength $J_{DW}$ in symmetric double wells. Under the DW Hamiltonian, the occupation imbalance operator evolves as

\begin{multline}
    \hat{n}_{l,r+1}(t) - \hat{n}_{l,r}(t) = [\hat{n}_{l,r+1}(0) - \hat{n}_{l,r}(0)]\cos{(2J_{DW}t/\hbar)}\\
    +i(\hat{a}_{l,r+1}^{\dagger}\hat{a}_{l,r}-\hat{a}_{l,r}^{\dagger}\hat{a}_{l,r+1})\sin{(2J_\mathrm{DW}t/\hbar)}.
\end{multline}

By choosing an evolution time of $\tilde{t} = h/(8J_\mathrm{DW})$ (i.e., a quarter rotation period $T/4$), the occupation imbalance is proportional to the current operator, $\hat{n}_{l,r+1}(\tilde{t}) - \hat{n}_{l,r}(\tilde{t}) = \hat{j}_{l,r}^{\parallel}/J$.

Similarly, to measure the current along the rung direction, we project the system onto isolated double wells along the rung direction, here denoted by the sites $(1,r)$ and $(2,r)$, and then apply the basis rotation with a complex-valued tunnel coupling $K_{DW}$. The density difference then evolves according to
\begin{multline}
    \hat{n}_{1,r}(t) - \hat{n}_{2,r}(t) = [\hat{n}_{1,r}(0) - \hat{n}_{2,r}(0)]\cos{(2K_\mathrm{DW}t/\hbar)}\\
    +i(e^{-ir\phi}\hat{a}_{1,r}^{\dagger}\hat{a}_{2,r}-e^{ir\phi}\hat{a}_{2,r}^{\dagger}\hat{a}_{1,r})\sin{(2K_\mathrm{DW}t/\hbar)}.
\end{multline}
At $\bar{t} = h/(8K_\mathrm{DW})$ (i.e., a quarter rotation period $T/4$), we can determine the rung current with single-bond resolution by measuring the number difference inside the double well, $\hat{n}_{1,r}(\bar{t})-\hat{n}_{2,r}(\bar{t}) = \hat{j}_{l,r}^{\perp}/K$.

Experimentally, we implement the basis rotations as follows: To measure the leg current, we project into isolated DWs along the leg direction by quickly increasing the long $y$-lattice to $35\,E_\mathrm{r,l}$ and changing the short $y$-lattice to $7.7\,E_\mathrm{r,s}$ in $\SI{150}{\micro\second}$. At the same time, we switch off the running-wave modulation to remove the coupling along the rung direction. Furthermore, we exchange the vertical confinement from a small-spacing vertical lattice to a large-spacing vertical lattice, reducing the vertical trap frequency from $2\pi\times\SI{4.7}{kHz}$ to $2\pi\times\SI{0.9}{kHz}$, and change the offset field to $\SI{17.0}{G}$, where the scattering length vanishes for the $\ket{3,3}$ state of cesium. Together, these steps initiate the DW dynamics with negligible on-site interactions. To stop the rotation after a quarter period, we increase the short $y$-lattice in $\SI{150}{\micro\second}$ to $37\,E_\mathrm{r,s}$, and immediately start recording the site-resolved occupations.

To apply the rung basis rotation, we cut the tunneling in leg direction by quickly increasing the short $y$-lattice in $\SI{150}{\micro\second}$ to $37\,E_\mathrm{r,s}$, and as for the leg rotation, changing the vertical confinement and the offset field. This initiates dynamics in the driven rung DW, with the modulation parameters kept unchanged. To halt the rotation, we abruptly switch off the running-wave modulation within \(\SI{150}{\micro\second}\) and image the occupation without any additional delay.

For both basis rotations, we need to independently calibrate the correct rotation duration (corresponding to a quarter period of the DW dynamics). As the DW dynamics correspond to simple $X$ rotations, we can calibrate them using (driven) DW tunneling oscillations. For this, we perform DW oscillations as described in Sect.~\ref{sec:sm_tunnel_calib} for the correct lattice parameters and fit the oscillation to extract the corresponding quarter-period evolution time. For the leg basis rotation, we obtain a typical tunnel coupling of $J_\mathrm{DW}/h=\SI{593(7)}{Hz}$, and for the rung basis rotation around $K_\mathrm{DW}/h=\SI{150(1)}{Hz}$. Note that the significantly higher frequency of the leg rotation makes it less susceptible to residual on-site interactions as well as local tilts $\Delta$, which modify the effective DW oscillation frequency according to $\sqrt{\Delta^2+4J_\mathrm{DW}^2}$. We have independently confirmed that the time obtained by DW oscillations is equal to the point where the measured current is maximal (corresponding to an exact rotation from density to current basis) when probing a state with known current and varying the basis rotation duration.

\begin{figure}[t]
    \centering
    \includegraphics[]{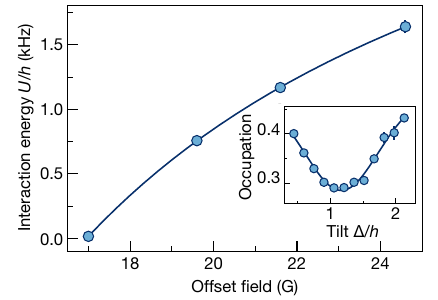}
    \caption{\textbf{On-site interaction energy calibration using tilt spectroscopy.} Calibrated on-site interaction energy $U/h$ as a function of the offset field. The solid curve is a fit to the experimental data using the equation for a magnetically tuned $s$-wave scattering length, $U(B)=U_0(1-\Delta_B/(B-B_0))$, where $U_0, \Delta_B$ and $B_0$ are free fit parameters. The error bars denote the standard-error-of-the-fit, and if invisible, are smaller than the marker size. The inset shows the average parity-projected site occupation as a function of tilt energy $\Delta$ at $\SI{21.6}{G}$, where the solid line is a fit of a Gaussian function to locate the tilt resonance position. The error bars denote the standard-error-of-the-mean over three repetitions, and if invisible, are smaller than the marker size. }
    \label{fig:U_calb}
\end{figure}

\section{Calibration}
\subsection{Energy offset}
The energy offset $\Delta$ between two adjacent lattice sites in a double well, which can be tuned by the superlattice phase and long-period lattice depth, is calibrated using modulation spectroscopy. In this process, one atom is initially loaded into the site with a lower potential of a tilted double well with energy offset $\Delta$. The system is then modulated using the running-wave lattice, which is also used for the laser-assisted tunneling scheme, at a given modulation frequency. The energy offset is determined by finding the resonant modulation frequency at which the average imbalance between double wells becomes zero.

\subsection{On-site interaction}
\textbf{Spectroscopic calibration}. -- We calibrate the on-site interaction energy $U$ by identifying the resonance where the tilt energy is equal to $U$~\cite{meinert_observation_2014}. After preparing a 1D Mott-insulating state with unity filling, we ramp the short lattice depth along both directions to the final values we used in the experiments. Subsequently, we introduce a staggered lattice potential $(-1)^i\Delta/2$, where $i$ is the site index, along the chain by ramping up the long-period lattice with a superlattice phase of $\pi/2$.

At $\Delta = U$, resonant tunneling to the nearest-neighbor sites is initiated, resulting in a coherent oscillation of doublons as $\lvert 1,1 \rangle_{\langle i,j \rangle} \leftrightarrow \lvert 2,0 \rangle_{\langle i,j \rangle}+\lvert 0,2 \rangle_{\langle i,j \rangle}$. By measuring the fraction of singly occupied sites after some evolution time, we can determine $U/h$ to be $\SI{1.64(5)}{kHz}$ at a magnetic field of $\SI{21.6}{G}$ (see the inset of Fig.~\ref{fig:U_calb}). We measure on-site interaction energies for different offset fields and find the scaling to be consistent with the estimation based on a magnetically tuned $s$-wave scattering length (Fig.~\ref{fig:U_calb}).

\begin{figure}[t]
    \centering
    \includegraphics[]{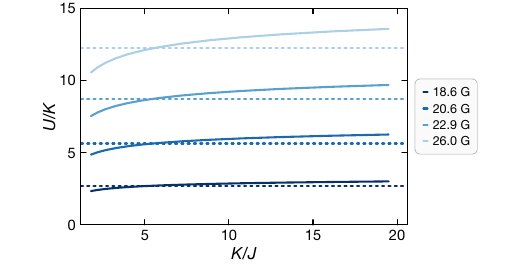}
    \caption{\textbf{Calibrated variation of $U$ with lattice depth.} For each offset field (i.e. scattering length), the calibrated interaction energy was extrapolated using a band structure calculation (solid lines). The dashed horizontal lines show the average $U$ for each offset field over the whole $K/J$ range.}
    \label{fig:sm_U_variation}
\end{figure}

\begin{figure*}[!t]
    \centering
    \includegraphics[]{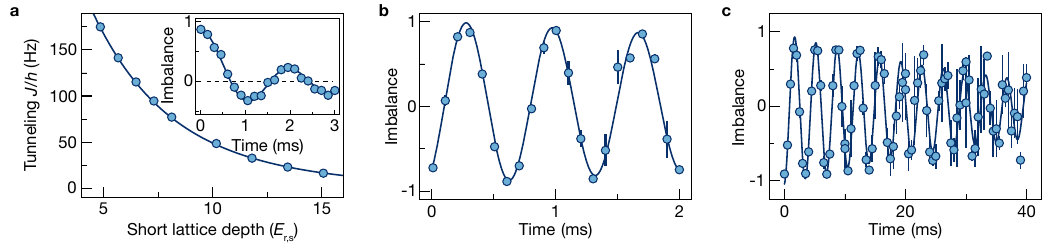}
    \caption{\textbf{Tunneling calibration.} \textbf{a,} Tunneling strength $J$ for various short-period lattice depths along the leg direction, calibrated via charge density wave (CDW) decay. The solid curve is a fit of an exponential function. The inset shows one example CDW decay trace for $6.5\,E_\mathrm{r,s}$, where the solid line is a fit of an exponentially decaying Bessel function to extract the frequency. \textbf{b,} Double well oscillations in a symmetric double well without drive. The data points show the time evolution of the imbalance at a short lattice depth of $9.5\,E_\mathrm{r,s}$ and long lattice depth of $50\,E_\mathrm{r,l}$. The solid line is a fit of an exponentially decaying sine function to the experimental data. \textbf{c,} Calibration of $K$ via driven double-well oscillations. Dynamics of the double well imbalance in the presence of the resonant running-wave modulation. The solid line is a fit of an exponentially decaying sine function, giving a driven tunnel coupling of $K/h=\SI{142.6(3)}{Hz}$ and a $1/e$ decay time of $\SI{42(7)}{ms}$. In all plots, the error bars denote the standard-error-of-the-mean whenever experimental data points are shown (two averages for each data point in a and b, three averages for each data point in c), or the standard-error-of-the-fit for fit results. If not visible, the error bar is smaller than the marker size.}
    \label{fig:tunn_calb}
\end{figure*}

\textbf{Band structure extrapolation}. -- The spectroscopic calibration of the interaction energy is done for one specific lattice depth. To infer the interaction energy for arbitrary lattice depths, we make use of the fully calibrated band structure (requiring knowledge of $z$ confinement, horizontal lattice depths and superlattice phase) and compute the interaction energy from the Wannier functions. The band structure extrapolation was confirmed to work well by comparison with an independently calibrated $U$ value for a different lattice depth.

In Fig.~\ref{fig:sm_U_variation}, this is shown for the interaction scaling experiment in Fig.~\ref{fig:meissner_ladders}d of the main text, where we tune $K/J$ via the short lattice depth along the leg direction. The same protocol is used to calibrate the change in $U$ when recording the bond currents as a function of $K/J$ in isolated plaquettes.

\subsection{Tunneling}
\label{sec:sm_tunnel_calib}
\textbf{Tunneling strength $J$ along the leg direction}. --  We calibrate the tunneling strength $J$ using the period-two charge density wave (CDW) dynamics~\cite{scherg_observing_2021,wienandEmergenceFluctuatingHydrodynamics2024}. After preparing the initial CDW using a superlattice along the leg direction, we switch off the long-period lattice and then abruptly decrease the short-period lattice to the final value. During the relaxation dynamics, we measure the time evolution of imbalance between odd and even sites $\mathcal{I} = (\langle \hat{n}_\mathrm{even}\rangle-\langle \hat{n}_\mathrm{odd}\rangle)/(\langle \hat{n}_\mathrm{even}\rangle+\langle \hat{n}_\mathrm{odd}\rangle)$. The tunneling strength along the leg direction is extracted by fitting an exponentially decaying Bessel function $\mathcal{I}(t) = A\mathcal{J}_0(4Jt/\hbar)\exp{(-t/\tau)}$, see inset of Fig~\ref{fig:tunn_calb}a. We calibrate the tunneling coupling for various short lattice depths and find good agreement of the scaling with a theoretical calculation of the band structure (Fig~\ref{fig:tunn_calb}a). Using this protocol, we also obtain an absolute calibration for the lattice depths of the short-spacing lattices in units of the recoil energy.

\textbf{Bare tunneling strength ${J_\mathrm{DW}}$ along the rung direction}. -- 
The tunnel coupling along the rung direction without laser-assisted tunneling is calibrated using imbalance oscillations in a symmetric double well. We prepare an initial state where one atom is occupied in the lower potential site of the double well using the superlattice phase. To initiate the tunneling, we suddenly set the superlattice phase to zero for a symmetric double well and change the short lattice depth in $\SI{150}{\micro\second}$ to its final value. We observe high contrast oscillations in the number difference between the two sites of the double wells with a frequency corresponding to $f=2J_\mathrm{DW}/h$ (Fig~\ref{fig:tunn_calb}b). 

\textbf{Driven tunneling strength along the rung direction $\mathbf{K}$}. --  The magnitude of the complex-valued tunnel coupling is calibrated using imbalance oscillations between double wells in the presence of laser-assisted tunneling. We prepare one atom in the double well with an energy offset $\Delta/h = \SI{4.7}{kHz}$ and initiate the driven tunneling by abruptly switching on the running-wave modulation. Fig~\ref{fig:tunn_calb}c shows single particle dynamics in a driven double well from which we extract the oscillation frequency. We find the tunneling strength $K/h$ to be around $\SI{150(1)}{Hz}$, which is in good agreement with the theoretical prediction $K = J_\mathrm{DW}\mathcal{J}_1(V_0/\hbar\omega)$.

\subsection{Running-wave parameters}

To restore tunneling in the rung direction using the running-wave modulation and obtain symmetric DWs in the effective, time-averaged model, the frequency detuning of the two beams that comprise the running-wave needs to fulfill the DW resonance condition, $\hbar\omega=\sqrt{\Delta^2+4J_\mathrm{DW}^2}$. Here, $\omega=\omega_1-\omega_2$ is the detuning between the two laser beams of the running-wave lattice, $\Delta$ is the energy difference between the two wells, and $J_\mathrm{DW}$ is the tunnel coupling in the DW. We use the first steps of the ground state preparation sequence to calibrate the resonance condition in a setting that closely resembles the actual experiments (Sect.~\ref{sec:sm_gs_prep}). Specifically, we start from a product state with one particle in a tilted DW, and then adiabatically turn on the running-wave modulation in $\SI{30}{ms}$ while simultaneously removing an extra tilt. This delocalizes the particle over each rung bond, and we can use the remaining imbalance as an observable to calibrate the resonance position. The atom becomes symmetrically delocalized across the DW with a vanishing imbalance only when the resonance condition is fulfilled. We experimentally verify that the imbalance remains constant over time, confirming the preparation of an eigenstate in the isolated DWs.

\begin{figure}[t]
    \centering
    \includegraphics[]{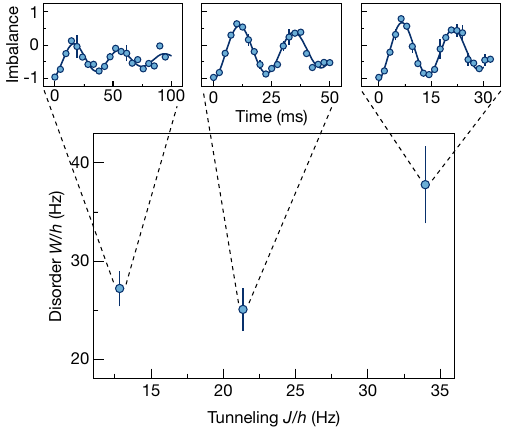}
    \caption{\textbf{Disorder estimation.} Estimation of the randomly distributed, white-noise on-site potential disorder from slow double well oscillations. From the steady-state imbalance at low coupling strengths (shown in the zoom-in plots, where the solid lines are fits of exponentially decaying sine functions, and the error bars denote the standard-error-of-the-mean over two repetitions), we can estimate the magnitude of potential disorder $W/h$. The error bars for the disorder are propagated from the standard-error-of-the-fit for the steady-state imbalance. The data is evaluated in a central region spanning $20\times20$ sites to avoid edge effects from the finite sharpness of the box walls.}
    \label{fig:dis_calb}
\end{figure}

\subsection{Flux}

As shown in earlier work by probing the running-wave phase pattern using current-current correlations in isolated double wells, the flux is almost exactly equal to $\pi/2$~\cite{impertroLocalReadoutControl2024}. However, this does not apply to the plaquette measurement, where the presence of the additional long lattice squeezes the two neighboring bonds in a plaquette, effectively reducing the flux. We calibrate the reduced flux in plaquettes using cyclotron orbits as demonstrated in~\cite{aidelsburgerExperimentalRealizationStrong2011}. In brief, we start with a single particle per plaquette, which is symmetrically delocalized across one leg bond. We then introduce a strong energy offset between the two legs (i.e., a DW tilt in the rung direction) and suddenly turn on the running-wave modulation to drive tunneling in the rung direction. Without an artificial gauge field, there would only be dynamics in the rung direction, as the symmetric superposition is an eigenstate of the symmetric DW in the leg direction. The gauge field creates an effective Lorenz-like force, bringing the particle into a circular motion around the plaquette. We fit this time evolution with an analytical expression for the single-particle dynamics and obtain a reduced plaquette flux of $0.71(2) \times \pi/2$ for the lattice parameters used in the plaquette measurement of Fig.~\ref{fig:plaquettes} of the main text.

\subsection{Disorder}
To estimate the average strength of random on-site potential disorder across our system, $W$, we use slow imbalance oscillations in double wells. We set a weak tunnel coupling in the double wells, $J_{DW}/h = 10\sim35\,\mathrm{Hz}$, which is on the same order as the overall disorder strength. We track the time evolution of the number imbalance between double wells and characterize the dynamics using an exponentially damped sine function, $\mathcal{I}(t) = \bar{\mathcal{I}}+\mathcal{I}_Ae^{-t/\tau}\sin{(\omega t+\phi)}$ where $\bar{\mathcal{I}}$ is the steady-state imbalance, $\mathcal{I}_{A}$ is the oscillation amplitude, $\omega = \sqrt{4J_\mathrm{DW}^2+W^2}/\hbar$ is the modified frequency due to the disorder, and $\phi$ is a phase to account for the finite ramp time (Fig~\ref{fig:dis_calb}). Assuming a white-noise disorder distribution within $[-W, W]$, we obtain
\begin{equation}
    W/h = \sqrt{\frac{6\bar{\mathcal{I}}/\mathcal{I}(0)}{1-\bar{\mathcal{I}}}}.
\end{equation}
Using this relation, we extract our potential disorder strength to be around ${W/h \approx \SI{30}{Hz}}$ from the steady-state imbalances.

\begin{figure}
    \centering
    \includegraphics[]{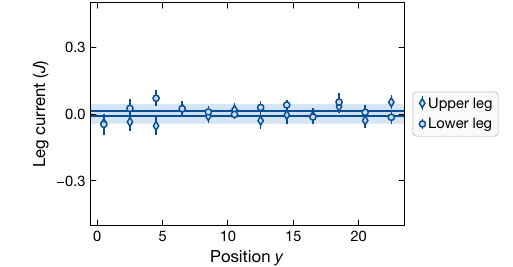}
    \caption{\textbf{Absence of currents in a ladder without flux.} Spatially resolved leg current measurement in a ladder without flux. The horizontal lines and the shading around it illustrate the spatial average current and its $1\sigma$-uncertainty across the ladder (same ROI as in the main text). We obtain an average of $-0.01(3)\,J$ for the upper leg and $0.01(3)\,J$ for the lower leg, both consistent with zero current. The error bars denote the standard-error-of-the-mean over 80 repetitions, and if invisible, are smaller than the marker size.}
    \label{fig:sm_bare_ladder_currents}
\end{figure}

\subsection{Reference current measurement in a ladder without flux}

We perform a reference current measurement for a state without any equilibrium currents to provide an additional benchmark for the current measurement. To this end, we prepare a ladder with only real-valued tunnel couplings in its ground state at half-filling. We start with the same initial state -- one particle per rung bond, which we then delocalize in a symmetric, bare double well at a rung tunnel coupling of around $J_\perp/h=\SI{165}{Hz}$. In a final step, the rungs are coupled together by increasing the leg tunnel coupling to $J/h=\SI{71(1)}{Hz}$, realizing similar parameters as the flux ladders with an interaction energy of around $U/h=\SI{890}{Hz}$. We then measure the leg currents with spatial resolution as before. As can be seen in Fig.~\ref{fig:sm_bare_ladder_currents}, we measure vanishing currents across all bonds of the ladder with a spatial average of $-0.01(3)\,J$ for the upper leg, and $0.01(3)\,J$ for the lower leg, respectively.

\section{Floquet micromotion and heating}

\subsection{Current lifetime}

\begin{figure}
    \centering
    \includegraphics[]{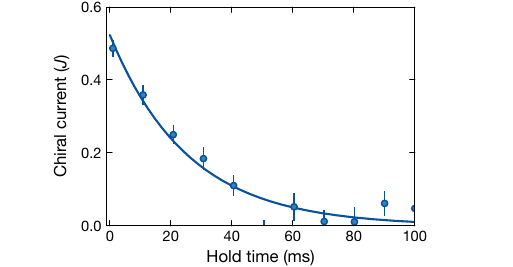}
    \caption{\textbf{Lifetime of the chiral current.} Lifetime of the chiral current in the Meissner phase for $K/J=2.19(3)$ and $U/J = 11.02(5)$. The solid line is an exponential fit, yielding a $1/e$-decay time of $\SI{26(3)}{ms}$. The error bars denote the standard-error-of-the-mean over roughly 30 repetitions, and if invisible, are smaller than the marker size.}
    \label{fig:sm_current_lifetime}
\end{figure}

The ground state phases exhibit finite lifetimes, which results in a decay of the equilibrium currents over time. This occurs both due to energy absorption from the periodic drive and from conventional technical heating, such as laser noise. To quantify this, we prepare a Meissner state using the rung coupling sequence and vary the hold time between state preparation and current measurement. Fig.~\ref{fig:sm_current_lifetime} shows the measured chiral current as a function of hold time for a Meissner state with $K/J=2.19(3)$ and $U/J = 11.02(5)$. The decay of the currents is well-described by a single exponential with a $1/e$-decay time of $\SI{26(3)}{ms}$. As the lifetime is on a similar order with typical ramp durations, this highlights the importance of choosing suitable parameter regimes where heating is suppressed.

\subsection{Floquet micromotion}

While quantities such as average currents and densities that we studied in the main text show good agreement with numerical estimates based on the effective, time-averaged Hamiltonian, the actual physical system follows the full Hamiltonian including the time-periodic drive. It is therefore expected that some observables will exhibit significant deviations from the simple effective model, most prominently due to the micromotion within one period of the drive.

\begin{figure}
    \centering
    \includegraphics[]{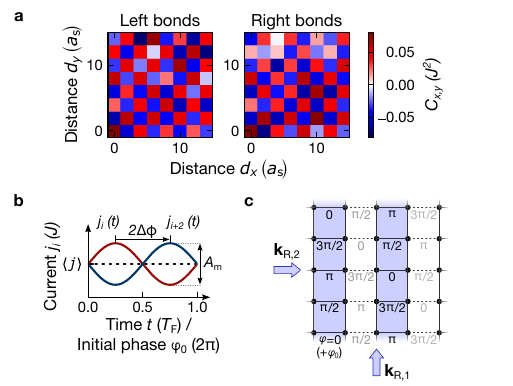}
    \caption{\textbf{Current-current correlations due to Floquet micromotion.} \textbf{a,} Current-current correlations for the plaquette measurement in Fig.~\ref{fig:plaquettes} of the main text, as an example for the vertical ('leg') bonds. Even though there should be no correlation between the same type of bond over different plaquette copies, the micromotion causes strong, long-ranged checkerboard correlations with an average magnitude of $0.04(2)\,J^2$. \textbf{b,} Schematic illustration of the evolution of the current operator within one Floquet period for two bonds with a distance of two short lattice sites. \textbf{c,} Phase distribution imprinted with the running-wave modulation, advancing by $\pi/2$ per bond.}
    \label{fig:sm_micromotion}
\end{figure}

\textbf{Correlation functions}. -- Experimentally, we have found striking deviations from the effective model when evaluating current-current correlation functions. To illustrate this, we take the plaquette current measurement of Fig.~\ref{fig:plaquettes} in the main text. Instead of evaluating average bond currents, we compute the connected current-current correlation function across distance, ${C^c_{x,y} = \langle\hat{j}_{i,j} \hat{j}_{i+d_x,j+d_y}\rangle - \langle\hat{j}_{i,j}\rangle\langle\hat{j}_{i+d_x,j+d_y}\rangle}$, as an example for the leg bonds. Starting from the effective model, all plaquettes should be identical, and hence, there should be no correlations between different plaquette copies for one bond type. However, in the experiment, we found strong checkerboard correlations when analyzing current correlations across the system (Fig.~\ref{fig:sm_micromotion}a). Their origin lies in the micromotion -- the difference between the effective Hamiltonian and the full model during one drive cycle. As illustrated in Fig.~\ref{fig:sm_micromotion}b, (local) observables oscillate around the average value of the effective model within one Floquet period. The strength of this modulation depends on the specific observable as well as the drive parameters. While this does not lead to spatial correlations by itself yet, the micromotion is additionally influenced by the spatial dependency of the periodic drive: To engineer a magnetic flux, every lattice site is modulated with a phase shift of $\pi/2$ w.r.t. its nearest neighbors (see phase distribution in Fig.~\ref{fig:sm_micromotion}c). This locally varying phase shift is also imprinted on the micromotion, causing the current on each bond to oscillate out-of-phase during a Floquet period. Since we probe at random times during the time evolution, the locally varying micromotion gives rise to strong spatial correlations that we are able to resolve. Specifically, the current correlations shown in Fig.~\ref{fig:sm_micromotion}a have an average magnitude of $0.04(2)\,J^2$, which is comparable to correlations expected from a physical signal with a strength of $0.2\,J$. This could impair the capability to detect current modulations that are non-stationary via correlation functions, such as for example expected in the vortex phase. In principle, it should, however, be possible to calibrate these artificial correlations by making use of the known phase distribution and, e.g., averaging over $\pi$-out-of-phase bonds before evaluating the correlation function. Note that while currents, being a measurement of local phases, are especially sensitive to micromotion, we also detected signatures of this in density correlations.

\subsection{Floquet heating and losses}

As commented on in the main text, a crucial aspect when working with interacting Floquet systems is to carefully determine parameter regimes where heating and atom losses are minimized~\cite{bilitewskiScatteringTheoryFloquetBloch2015,sun_optimal_2020}. During our study, we have identified a few key contributions:

A significant heating channel arises from driving band resonances along the leg direction via the running-wave modulation. Among those, excitations from the first to the third Bloch band can be effectively driven, both at the fundamental frequency $\omega_b^{(3)}-\omega_b^{(1)}$ as well as half the frequency $(\omega_b^{(3)}-\omega_b^{(1)})/2$ via a two-photon process. For typical lattice depths around $8\,E_\mathrm{r,s}$, the fundamental frequency is at around $\omega_3-\omega_1=2\pi\times\SI{10.5}{kHz}$. Even with moderate modulation depths, the resonance can be more than $2\pi\times\SI{2}{kHz}$ wide as observed from atom-loss spectroscopy, where we detect the remaining atom number after typical experiment durations. The two-photon process at half the frequency is still around $2\pi\times\SI{500}{Hz}$ wide. The width and heating rate increase both with the modulation depth and the tunnel coupling $J$ (via the bandwidth).

While the former processes also exist in a single-particle system, the interaction energy in a many-body system opens up additional heating channels. In particular, when the modulation frequency coincides with the on-site interaction energy (or twice as well as fractions thereof), multiply-occupied sites can be resonantly created or depleted, which can be lost via two- or three-body loss processes. Additionally, driving this process leads to further unwanted terms in the effective Hamiltonian. 

We experimentally probe the different loss channels spectroscopically, using one-dimensional chains as a controlled reference system. For a typical tunnel coupling on the order of $J/h=\SI{100}{Hz}$, we use interaction energies up to $U/h = 1-1.5\,\mathrm{kHz}$ to reach the strongly correlated regime. As a result, we found the entire frequency range up to a modulation frequency of \(\SI{3}{kHz}\) to be too lossy for practical use.

In summary, these effects set a tight constraint on the usable parameter regimes as well as preparation sequences, in particular since they are all on the same order of magnitude of $1-10\,\mathrm{kHz}$. For our specific system, defined by the combination of atomic species and lattice spacings, the remaining usable modulation frequency range is $4\text{–}5\,\mathrm{kHz}$, where we observe negligible losses over the timescales of state preparation. This also highlights the advantages of our superlattice-based scheme compared to for example tilted lattices based on a magnetic gradient: Firstly, achieving such large tilts at the single-site level using magnetic gradients is highly challenging, while an array of tilted DWs or a staggered superlattice straightforwardly realizes tilts on the $1-10\,\mathrm{kHz}$ level. Secondly, in a tilted lattice long-range tunneling events can become resonant, which sets yet another constraint on the usable modulation frequencies.

\section{Data analysis}

The raw fluorescence images are first converted into site-resolved, parity-projected occupations using a reconstruction algorithm~\cite{impertro_unsupervised_2023}. In a next step, we filter out experimental realizations where the overall imbalance in an image is larger than $0.2$, which originates from short-term fluctuations or environment-related drifts of the superlattice phase lock (all experiments are conducted at zero overall imbalance, fraction of discarded images usually $\lesssim 5\,\%$). For data analysis, it is crucial to accurately identify which sites belong to a ladder and which belong to a double well for the current measurement. To determine this, we compare the current-current correlations for the two possible partitions, as shown in Fig.~\ref{fig:sm_micromotion}. For the correct partitioning, we observe alternating positive-negative correlations, while the wrong partitioning leads to positive low-distance correlations. At this point, the currents can immediately be read out by computing the left-right population difference without further processing being applied.

\section{Numerical simulation}

To benchmark the experimental measurements, we perform different numerical studies. For small systems such as plaquettes and non-interacting systems, we employ exact diagonalization (ED), while for larger and interacting systems, we make use of tensor network algorithms.

\subsection{Exact diagonalization}

Simulations based on exact diagonalization are implemented using the python package \texttt{QuSpin}~\cite{weinberg_quspin_2019}. Based on the resulting ground-state wave function, we can directly evaluate the expectation values of current or density operators as well as correlators.

Using this method, we simulate the two-particle plaquette ground states, ladders in the non-interacting limit, as well as the small ladder systems for estimating the effective temperature. In all simulations, we take into account parity projection by evaluating the observables in a parity-projected Fock basis with all on-site occupations taken $\mod 2$, as detailed also below.

\subsection{DMRG}

Simulations based on tensor networks are performed using the python package \texttt{TeNPy}~\cite{hauschild_efficient_2018}. All ground-state wave functions are computed with the density matrix renormalization group (DMRG) algorithm. We typically use a local cutoff of $n_\mathrm{max}=3$ bosons per site to limit computational cost in the interacting regime and truncate at a bond dimension between $\chi=600...1024$ depending on the specific parameters, verifying for different $\chi$ that the algorithm has indeed converged. The returned ground-state wave function directly gives us access to the ideal expectation values of density and current operators, as well as correlation functions. 

To simulate the effect of parity projection, we take the computed ground-state wave function and evolve it in isolated double wells using the time-dependent variational principle (TDVP) algorithm for a quarter period, exactly as it is done in the experiment. This has transformed from current to density basis, and we can now sample density snapshots using \texttt{TeNPy}'s built-in sampling algorithm. From the left-right density difference, we obtain the local particle current, which can be checked to coincide with the expectation value of the current operator as evaluated directly from the ground-state wave function. Parity projection is now straightforwardly added by taking the sampled occupations $\mathrm{mod}\,2$ before evaluating the current. We usually sample around $10^4$ snapshots to have sufficient statistics.

For the Meissner current scaling (Fig.~\ref{fig:meissner_ladders}d of the main text), we used a chain length of $L=20$ and a bond dimension of $\chi=600$. For the chiral current phase diagram (Fig.~\ref{fig:ladder_phase_diagram}a of the main text), we used a chain length of $L=64$ and a bond dimension of $\chi=1024$.

The many-body gap diagrams in Fig.~\ref{fig:ladder_prep} and Fig.~\ref{fig:ladder_phase_diagram}a of the main text are also computed using the DMRG algorithm ($L=20$ sites), where we approximate the energy of the first excited state by orthogonalizing against the ground state. 

\subsection{Temperature estimation}
To estimate the effective temperature of the system as described by the effective Hamiltonian, for our case, the ladder-HBH-model, we perform small system exact diagonalization using the python package \texttt{QuSpin}~\cite{weinberg_quspin_2019} and evaluate the finite temperature expectation value for density correlations~\cite{kaufman_quantum_2016,spar_realization_2022}. With the assumption that many-body statistics of the system are described by the canonical ensemble, the expectation value for an observable $A$ at a finite temperature $T$ is calculated as
\begin{equation}
    \langle A \rangle = \sum_{\Omega}e^{-E_\Omega/k_BT}\langle \psi_\Omega \rvert A \lvert \psi_\Omega \rangle
\end{equation}
where $\psi_{\Omega}$ denotes the eigenstate, which is labelled by $\Omega$, $E_{\Omega}$ is the corresponding eigenenergy, and $k_B$ denotes the Boltzmann constant. For each $K/J$ and $U/J$, we obtain a full eigenspectrum in the Fock space basis considering up to three bosons per site and six particles in 2$\times$6 ladder system. Then, the observable $A$ is given by
\begin{equation}
    \langle A \rangle = \sum_{\Omega} e^{-E_\Omega/k_BT}\sum_{j} \lvert c_{\Omega,j} \rvert^2 \langle \cdots, n_k, \cdots \rvert A \lvert \cdots, n_k, \cdots \rangle ,
\end{equation}
with each eigenstate $\lvert \psi_{\Omega} \rangle = \sum_j c_{\Omega,j}\lvert \cdots,n_k,\cdots \rangle$ being constrained to a fixed particle number $\sum_kn_k = 6$. We evaluate the connected rung correlator including parity projection, $\langle \mathcal{P}_{1,r}\mathcal{P}_{2,r} \rangle - \langle \mathcal{P}_{1,r} \rangle \langle \mathcal{P}_{2,r} \rangle$, where $\mathcal{P}_{l,r}$ is the parity-projected particle number operator for the site $l=1,2$ of the $r$-th rung. 

To compare with the experimentally measured density correlations, we simulate the system using the calibrated Hubbard parameters for both five and six particles, including parity projection as described above, at several different temperatures. We then average the result for the two particle numbers, five and six, which allows us to predict the values for our experimental filling of around $0.45$. As shown in Fig.~\ref{fig:ladder_phase_diagram}c of the main text, the measured correlation magnitude in the range of $K/J=1 \ldots 2$ is compatible with a temperature on the order of $k_BT \sim J$. Towards the vortex regime, the temperature is likely higher due to the smaller many-body gap.

\begin{figure}
    \centering
    \includegraphics[]{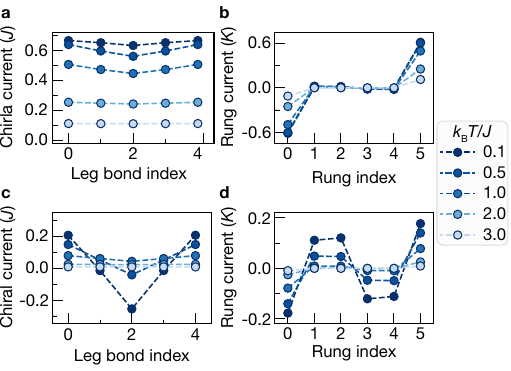}
    \caption{\textbf{Finite temperature simulation for Meissner and vortex phase.} Small-scale ED simulation, showing the chiral leg current and rung current distributions for $K/J = 2$ (Meissner phase, \textbf{a} and \textbf{b}) and $K/J=0.5$ (vortex phase, \textbf{c} and \textbf{d}) for different temperatures.}
    \label{fig:sm_finiteT_sim}
\end{figure}

Additionally, we investigate the temperature dependence of the current distribution using the same finite temperature numerical simulation as described above. We set $U/J = 11$ and $K/J =$ 2 and 0.5, which correspond to the Meissner and vortex regime, respectively. In the Meissner regime, at $k_B T = J$, the current distribution retains the characteristic homogeneous profile but with a reduced current of $\sim70\,\%$, which is at a similar scale as the strength measured experimentally. In contrast, the modulation amplitude of the current distribution in the vortex regime almost fully vanishes at $k_BT = J$, making it experimentally much harder to detect. Furthermore, at a finite temperature, the location of the vortices might not be pinned by the boundaries, requiring one to evaluate correlation functions instead of average currents. 

\begin{center}
\vspace*{0.5em}
\textbf{SUPPLEMENTARY REFERENCES}
\end{center}
\vspace{0.5em}

\putbib[manuscript]
\end{bibunit}